\documentclass[%
 superscriptaddress,
 amsmath,amssymb,amsfonts,
 aps,
 prb,
 twocolumn, 10pt,
 longbibliography,
 floatfix
]{revtex4-1}
\usepackage[utf8]{inputenc}
\usepackage{graphicx}
\usepackage[sf]{subfigure}
\usepackage{color}
\usepackage{fouriernc}
\usepackage[dvipsnames,svgnames,x11names,hyperref]{xcolor}
\usepackage[low-sup]{subdepth}
\usepackage{MnSymbol}
\usepackage{multirow}
\usepackage{outlines}
\usepackage{etoolbox}
\usepackage{enumitem}
\bibpunct{[}{]}{,}{n}{}{}

\definecolor{linkcolor}{RGB}{6,69,173}
\usepackage[T1]{fontenc}
\usepackage[utf8]{inputenc}
\usepackage[colorlinks=true,
            linkcolor=linkcolor,
            urlcolor=linkcolor,
            citecolor=linkcolor,
            unicode,
            pdfencoding=auto]{hyperref}

\usepackage[
]{physpack}

\begin{document}

\title{Entanglement Dynamics between Ising Spins and a Central Ancilla}
\author{Joseph C. Szabo}
\affiliation{Department of Physics, The Ohio State University, Columbus, Ohio 43210, USA}

\author{Nandini Trivedi}
\affiliation{Department of Physics, The Ohio State University, Columbus, Ohio 43210, USA}

\begin{abstract}
We investigate competing entanglement dynamics in an open Ising-spin chain coupled to an external central ancilla qudit. In studying the real-time behavior following a quench from an unentangled spin-ancilla state, we find that the ancilla entanglement entropy $S_{vN;\mathcal{A}}$ tracks the dynamical phase transition in the underlying spin system. In this composite setting, purely spin-spin entanglement metrics such as mutual information and quantum Fisher information (QFI) decay as the ancilla entanglement entropy grows. We define multipartite entanglement loss (MEL) as the difference between collective magnetic fluctuations and QFI, which is zero in the pure spin chain limit. MEL directly quantifies the ancilla's effect on the development of spin-spin entanglement. One of our central results is that  $MEL(t) \propto e^{S_{vN;\mathcal{A}}(t)}$. Our results provide a platform for exploring composite system entanglement dynamics and suggest that MEL serves as a quantitative estimate of information entropy shared between collective spins and the ancilla qudit. Our results present a new framework that connects physical spin-fluctuations, QFI, and bipartite entanglement entropy between collective quantum systems.
\end{abstract}
\date{\today}

\maketitle

\section{Introduction}
Entanglement is the most fascinating and perplexing feature of composite, many-particle quantum systems. Understanding its origin in physical platforms; whether due to particle statistics, correlations, and/or dynamical interactions, lies at the heart of quantum matter, sensing, and algorithm development ~\cite{Vidal_2003, Giovannetti_2011, preskill_2012}. This pursuit is being driven by technological advances in cold-atomic condensates~\cite{Sorensen2001, esteve2008,  Strobel14, Bernien_2017, Guardado_2018}, trapped-ion platforms~\cite{Blatt2008, Kim2010, Monroe2013, Jurcevic2014, Jurcevic2017, Bollinger2016, Garttner2017, hempel2018, monroe2019programmable, Tan2021}, cavity QED~\cite{Pellizzari1995, Leroux_2010, krischek2011, Davis_2019}, and superconducting circuits~\cite{Arute2019, satzinger2021}, which have brought dynamical quantum systems to the forefront of experimental and theoretical research. These experiments allow us to study how entanglement develops from purely classical initial states in the presence of quenched interactions, dissipation and driving, and decoherence. Here, many fundamental questions regarding entanglement remain unanswered such as determining it's relationship to physical properties as well as understanding the connections among the various methods for measuring entanglement.

To quantify entanglement, we work along two fronts: (i) through a partitioning structure that measures entanglement entropy, providing an information theoretic characterization of the number of entangled degrees of freedom (DoF) shared between subsystems; and (ii) through holistic multipartite entanglement measures that capture collective quantum correlations. 
Entanglement entropies have proven to be an essential theoretical diagnostic for characterizing quantum phases~\cite{nielsen00, Hastings_2007}, detecting topological order~\cite{kitaev2006, tee_wen}, and understanding nonequilibrium quantum dynamics and thermalization in pure quantum systems~\cite{kim2013, vosk2015, Altman2018}. From the experimental side, entanglement entropy remains one of the most challenging measures, as entropy does not remain an entanglement monotone when subject to loss, dissipation, and interaction with an environment. Few or multi-particle correlators accessible to realistic experimental platforms can only be related to entropy in fine-tuned integrable models~\cite{fermionic_entanglement_Zandardi, Peschel_2009, Swingle2010}, otherwise entropy calculations require full-state tomography, robust multi-correlations measures that provide a lower bound estimate~\cite{Brydges2019, Imai2021}, or many-body interference~\cite{Islam2015, Lukin2019}. On the other hand, multipartite entanglement witnessed through Quantum Fisher information (QFI) provides a true entanglement monotone regardless of classical entropy contributions, and in pure quantum systems can be measured through collective single particle operators~\cite{Hyllus_2012, Strobel14}. In thermal equilibrium, QFI is directly related to the dynamical response function of few particle operators~\cite{QFI_dynamical_susceptibility_Zoller, pontus2021, Scheie2021}. Though entropy and QFI provide qualitatively different perspectives on entanglement, they are essential parts of the same phenomenology: phase transitions, quantum thermalization, many-body localization.

QFI and entropy have been studied extensively with regards to quantum information dynamics in nonequilibrium and open environments. Each provide a unique perspective regarding the transition from semiclassical to quantum behavior as well as establishing myriad thermalization physics of quantum systems. Despite their popular theoretical investigations, few direct connections between the two have been established. 

Recent theoretical works have revealed qualitative relations connecting correlations, entanglement entropy, and QFI in a purely semiclassical perspective for collective spin and light-matter systems~\cite{lerose2020}. In such systems, it is well-known that collective light-matter entanglement is necessary for generating effective long-range spin Hamiltonians and achieving highly squeezed, multipartite entangled spin states~\cite{Ma_2009, Leroux_2010, Gietka_2019, Davis_2019, lewis-swan2019}. Though we have this heuristic understanding, no numerical connection between QFI and entropy has been interrogated. Theoretical studies examining the fate of entanglement subject to classical entropy, suggest that strong connections between QFI and entropy exist and their dynamics are intimately related~\cite{brenes2020, wybo_bath2020, Alba_2020, maity2020}. Beyond these, few works directly investigate the dynamical connection between entropy and multipartite entanglement precisely in fully quantum systems away from integrability and semiclassical approximations. The questions that remain to be answered are how do composite systems share quantum information, what are the implications for subsystem entanglement content, and what are the roles of various quantum information metrics?

In this paper, we provide general insights with regards to these questions by studying local quantum systems beyond integrability. We use numerically exact methods to study the entanglement dynamics of a quenched Ising spin-chain with weak exchange with a multilevel ancilla qudit. The Ising model provides a highly studied, integrable starting point for understanding nonequilibrium entanglement dynamics, where entropy and QFI have been thoroughly characterized in various limits. The ancilla then provides an environment through which the spin ensemble can undergo loss and decoherence while at the same time allowing us to interrogate the ancilla degrees of freedom (DoF) as it relates to the reduced, spin-subsystem. Interestingly, such a central mode relates to cavity quantum electrodynamics (cQED) experiments~\cite{Naik_2017, Pechal_2018}, central spin or dressed cavity in cold-atoms ~\cite{Zhang_2010, Zeiher_2016, Saffman_2009, Jau_2016} or trapped ions~\cite{Figgatt_2017}, nuclear magnetic resonance (NMR), and is a common element of quantum computing codes employing an oracle vertex in a qubit network~\cite{Grover_1997, Figgatt_2017, anikeeva_2020}. In what follows, we briefly outline the key results of our numerical investigation, review the entanglement metrics employed in this paper, and finely discuss the microscopic Hamiltonian and resulting dynamics.


\begin{figure}
    \centering
    \includegraphics[width=0.45\textwidth]{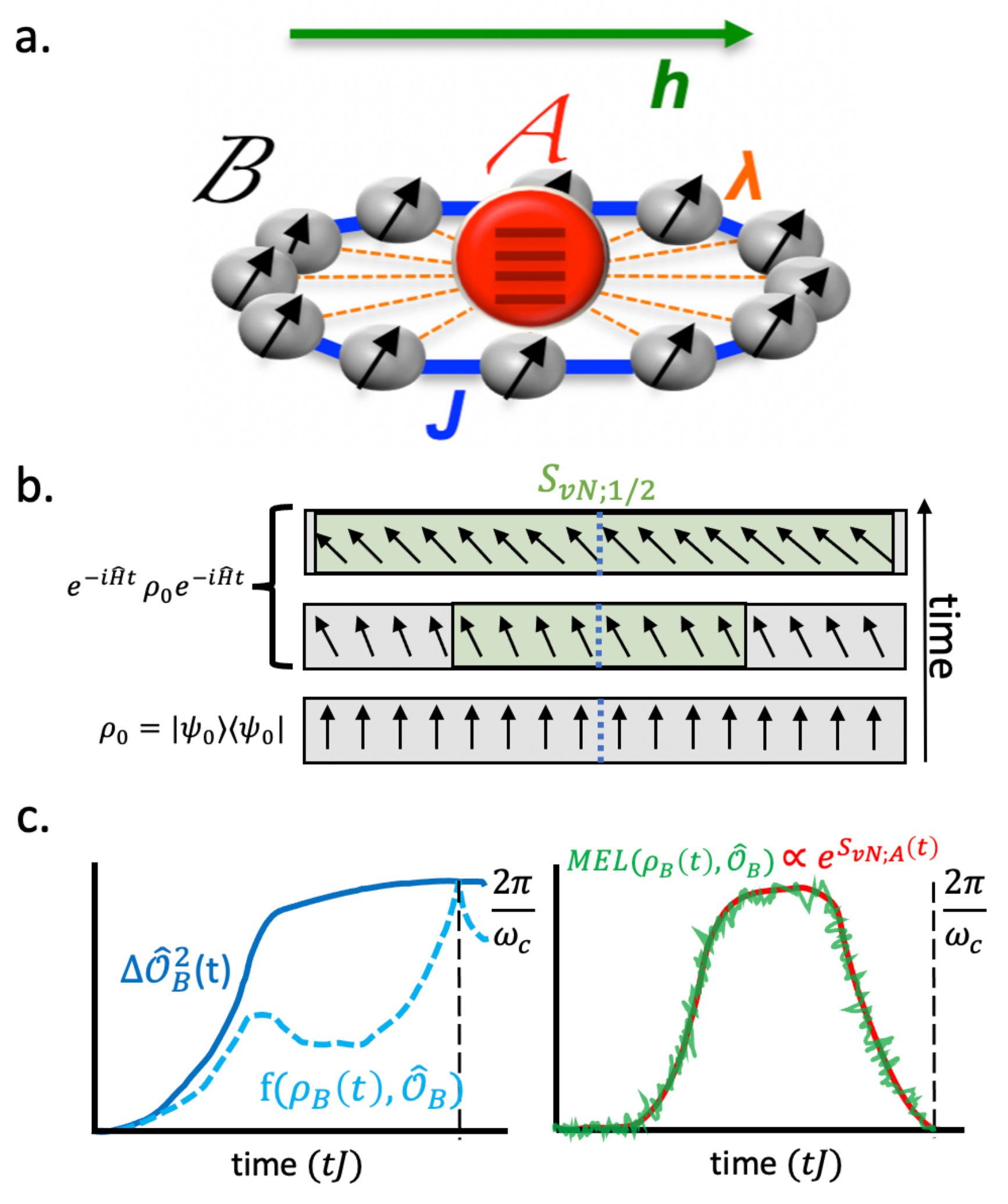}
    \caption{Spin-ancilla schematic with $\mathcal{B}$ and $\mathcal{A}$ representing the 1D spin-chain and multi-level ancilla, respectively. Ising spin interactions given by $J$, magnetic-field $h$, and spin-ancilla coupling $\lambda$. (b) Cartoon representation of the dynamical quench from an initially polarized spin state. Entangled region across the center boundary (green) grows in time as quasiparticles within a distance $t / v_B$ are transmitted between the two bipartite regions. (c) Schematic representation of our key finding in the interacting spin-boson composite system; operator fluctuations $\Delta\hat{\mathcal{O}}_B(t)$ and the corresponding quantum Fisher information $f(\rho_B(t), \hat{\mathcal{O}}_B)$ within the spin subsystem begin to deviate in time as external entropy $S_{vN;\mathcal{A}}$ grows (red). This discrepancy defined at $MEL$ (green) develops proportionally to the $e^{S_{vN;\mathcal{A}}(t)}$. The auxiliary bosonic system entanglement oscillates like the characteristic level spacing $\omega_c$.}
    \label{fig:cQudit}
\end{figure}

\section{Summary}

In this paper we analyze the subsystem entanglement dynamics of a quenched Ising spin-system with each site uniformly coupled to a qudit ancilla. We specifically focus on 
\begin{enumerate}[label=\roman*]
    \item Ancilla entanglement: $S_{vN;\mathcal{A}}$
    \item Mutual information: $MI = S_{vN;1/2} - \frac{1}{2}S_{vN;\mathcal{A}}$
    \item QFI $(F)$ and spin fluctuations: $F(\rho, \hat{S}_\mu) \leq \Delta  \hat{S}_\mu^2$
\end{enumerate}
This work represents one of the first studies characterizing the fate of entanglement in quenched quantum matter coupled to a controllable environment, here a bosonic mode. We work in the large bosonic-limit where results are converged with respect to the bosonic dimension ($q >> L$). Without any interaction between spins and ancilla, we observe the dynamical quantum phase transition associated with GGE description of the interacting spins and the transition from area to volume-scaling $1/2$-chain entanglement entropy, $S_{vN} \sim \mathcal{O}(L)$, where $L$ is the number of spins in the 1D-chain. We similarly find that long-time saturation value of the transverse and longitudinal spin-fluctuations agree with previous analytic results.

We find surprising physics away from integrability in the full spin-ancilla model, which admits simple intuitive extensions from the isolated Ising spin chain limit. For weak-coupling between spins and ancilla we observe:

\begin{enumerate}[label=\roman*]
    \item Ancilla entanglement entropy $S_{vN;\mathcal{A}}$ tracks the underlying spin DQPT, which persists under nonlocal ancilla coupling.
    \item $S_{vN;\mathcal{A}}$ scales like $\propto \log{L}$.
    \item Mutual Information (MI) and QFI within spins decreases monotonically with coupling as expected between system-environment.
    \item We define this discrepancy between the variance in spin-spin correlations (fluctuations) and spin-spin multipartite entanglement (QFI) as multipartite entanglement loss (MEL).
    \item Entanglement loss witnessed by MEL$(t)$ develops proportionally to $e^{S_{vN;\mathcal{A}}(t)}$.
\end{enumerate}
Our most significant finding is that the corresponding growth and decay of entanglement within the ancilla and spins respectively are intimately connected. We find that the discrepancy between spin-spin fluctuations and QFI captures the entanglement profile of the ancilla and provides a strong estimation on information transfer. This formulation comes from a complementary approach to previous results, where we find an identical log relationship that connects correlations and entanglement~\cite{lerose2020}. This relationship not only provides a strong estimate in the long-time limit, but accurately captures the magnitude of real-time dynamics. We expect that this small, finite ancilla behavior captures essentially semiclassical entanglement features of the collective interacting spin-system and provides an exciting frame to understand thermalization of many-body systems with their environment.


\section{Entanglement Metrics} 
The key measures of entanglement in this work are entanglement entropies (von Neumann and mutual information) and multipartite entanglement (spin-fluctuations and quantum Fisher information). The former two rely on bipartitions of the Hilbert space to calculate entropy, while the latter characterizes holistic entanglement content shared between eigenvalues of collective few-body operators. 

The von Neumann entanglement entropy taken between two regions ($\mathcal{A},\mathcal{B}$) is calculated as

\begin{equation}
    S_{vN} = \sum_k \lambda_k \log(\lambda_k).
\end{equation}
Where $\lambda_k$ are eigenvalues of the reduced density matrix (RDM) obtained by integrating out either subsystem $\mathcal{A} \text{ or } \mathcal{B}$. Though insightful in theoretical investigations, $S_{vN}$ is not an entanglement monotone for open quantum systems. In this scenario where we employ a spin-chain and an ancilla, the ancilla serves as an environment when considering the entropy in the spin-chain. Therefore, we calculate the $\frac{1}{2}$-chain mutual information (MI). For a 3-body example defined by total Hilbert space $\mathcal{H} = \mathcal{H}_a \otimes \mathcal{H}_b \otimes \mathcal{H}_c$ the mutual information between $ab$ is
\begin{equation}
    \mathcal{I}(ab;c) = S_{vN}(a|bc) + S_{vN}(b|ac) - S_{vN}(ab|c),
\end{equation}
where $S_{vN}(i|j)$ is the standard bipartite entanglement entropy of reduced space $i$, explicitly tracing out $j$. Mutual information serves as a useful rectification to entanglement entropies' failure in open quantum systems and is the subject of recent theoretical investigations~\cite{maity2020}.

QFI on the other hand, is an entanglement monotone regardless of purity and environment~\cite{Helstrom_1969, Toth_2014, Pezze_2018}. In a pure quantum state it is precisely equal to collective fluctuations witnessed by operator $\hat{\mathcal{O}}$

\begin{equation}
    \mathcal{F}(\hat{\mathcal{O}}, \rho) = 2\sum_{\alpha,\beta}\frac{v_\alpha - v_\beta}{v_\alpha + v_\beta}|\langle u_\alpha |\hat{\mathcal{O}}|u_\beta\rangle|^2 \leq 4\langle \Delta\hat{\mathcal{O}}^2 \rangle\  \cdot
    \label{eq:pQFI}
\end{equation}
$u_\alpha, v_\alpha$ are the eigenvectors and eigenvalues of the density matrix $\rho$. QFI originates from metrological studies and refers to the precision of a measurement conditioned on a global operator $\hat{\mathcal{O}}$ and state $\rho$. Given a unitary transformation of the form 
$ 
    \hat{U} = e^{i\hat{\mathcal{O}}\theta}$ and $ 
    \rho_\theta = U\rho U^\dagger$,
the precision in parameter $\theta$ is constrained by the quantum Cramer-Rao bound $
    \Delta\theta^2 = \frac{1}{\mathcal{F}(\hat{\mathcal{O}}, \rho)} $ .
For pure states, the uncertainty in $\theta$ is related to the uncertainty in its conjugate operator. For spin systems and operator defined as $\hat{\mathcal{O}} = \sum_i \vec{s}_i\cdot\hat{n}_i$, QFI is identical to the spin fluctuations about the vector $\vec{S}^n$ on the Bloch sphere. For mixed states, $\mathcal{F}(\hat{\mathcal{O}}, \rho)$ must be written in terms of an eigendecomposition of the initial mixed density matrix. In general, different operators will present different bounds on Eq.\ref{eq:pQFI}, and there has been no analytic insight into determining how this bound can be made tighter in arbitrary quantum systems away from purity.

Beyond metrological optimization, QFI witnesses multiparticle entanglement within a state $\rho$ and even more recently in the detection of topological quantum phases \cite{Pezze2017, Lambert2020}. The Fisher information density $f(\hat{\mathcal{O}}, \rho) \equiv \mathcal{F}(\hat{\mathcal{O}}, \rho)/N$ given $N$ constituent particles and $\hat{\mathcal{O}}$ being a sum over local site-operators provides that for $f > k$ the system is at least $(k+1)$-partite entangled. Determining the true multiparticle entanglement of a system requires optimization over the set of all possible $\hat{\mathcal{O}}$.

Though, QFI depends on the choice of collective operator $\hat{\mathcal{O}}$, general relationships exist that connect the development of multipartite and bipartite entanglement. Entropy provides a logarithmic measure of how spread a quantum state is throughout the full Hilbert space, while multipartite entanglement examines the development of off-diagonal components of the density matrix. For unentangled initial states, entanglement entropy generically grows proportionally to $t$ unless subjected to localization physics, integrability, or conservation laws, while QFI and spin-fluctuations grow like $e^{\alpha t}$ with some phenomenological exponent $\alpha$. In quadratic bosonic or fermionic systems this heuristic relation is exactly proportional; $S_{vN}(t) \propto \log{\mathcal{F}(\hat{\mathcal{O}}_\text{max}, \rho(t)})$, where $\hat{\mathcal{O}}_\text{max}$ is a generic operator that witnesses the maximal fluctuations at each instance in time. This definition relies on translational invariance and a homogeneous collective system~\cite{Hackl2018, Bianchi2018, lerose2020}. Beyond such systems, for arbitrary Hermitian matrices, the equality in Eq.\ref{eq:pQFI}, is believed to similarly develop like $\sim e^{S_{vN;\mathcal{B}}}$, but has only been characterized in random matrices with Hilbert spaces $\mathcal{O}(2 - 10^2)$~\cite{toth2018lower}. Beyond these conditions no analyses have been performed to relate bipartite entropy and QFI with regards to general composite quantum systems and physical, well-characterized operator dynamics.

In this work we consider the simplest realization of an interacting composite system. We study an interacting spin-ancilla system to illuminate how observables and information in interacting quantum spins behave under exchange with a bosonic ancilla. To capture the difference between fluctuations and true multiparticle entanglement defined in Eq.\ref{eq:pQFI} over the reduced spin density matrix, we define the multipartite entanglement loss (MEL) as

\begin{equation}
    MEL(\hat{\mathcal{O}}, \rho) = \langle \Delta\hat{\mathcal{O}}^2 \rangle - \frac{1}{4}\mathcal{F}(\hat{\mathcal{O}}, \rho_{\mathcal{B}}).
\end{equation}
This counts the difference in fluctuating constituent particles from the internal contribution to the multiparticle entanglement, such that for a pure system returns zero missing entanglement. The operator $\hat{\mathcal{O}}$ is specifically conditioned on the total Hilbert subspace spanning the system and ancilla $\hat{\mathcal{O}} \sim \hat{\mathcal{B}} \otimes \hat{\mathcal{A}}$, while in defining QFI, we trace over the ancilla degrees of freedom in both $\hat{\mathcal{O}}$ and $\rho$. We specifically focus on $\hat{S}_n \otimes \hat{\mathcal{I}}$ for collective spin magnetization in the system and identity in the ancilla.

Provided we are working with an essentially free-fermion model (JW transformation of the 1D Ising model) and a quadratic fermion-boson coupling, it would be interesting to observe a similar relationship between fluctuations and entanglement entropy as mentioned above. 

Our focus in this paper is understanding how the inequality in Eq.\ref{eq:pQFI} (MEL) evolves in a specific model of an interacting spin system coupled to an ancilla described below. We observe for the first time a strong quantitative connection between MEL and $S_{vN : \mathcal{A}}$:

\begin{equation}
    MEL(\rho, \hat{S}_{max}) \propto e^{S_{vN,\mathcal{A}}} - 1.
    \label{eq:result}
\end{equation}
$\hat{S}_{max}$ refers to the maximal MEL optimized over collective spin observables. Here we cannot perform such an extensive search over operators but find very strong agreement with simple collective operators in a well-characterized model. In the pure spin system limit, there is no discrepancy between the spin fluctuations and the corresponding QFI measured over the reduced density matrix of the spin sector. Here $MEL = 0$, and the entanglement entropy with the system must similarly be $0$. In the opposing limit where spin fluctuations $\Delta S_n^2 \propto L^2$ and $\mathcal{F}(\hat{S}_n, \rho_{\mathcal{B}}) = 0$, this restricts the ancilla entanglement to be at most $\mathcal{O}(\log L)$. We expect that Eq.\ref{eq:result} is only accurate when the environment experiences an effective semiclassical system, or in other words is coupled to collective degrees of freedom to the quantum system of interest. In a completely locally thermalizing regime, the entropy of the quantum system should scale like $\mathcal{O}(L)$, so observing this semiclassical result in a local quantum system is surprising.



\medskip

\section{Model}
The transverse-field Ising chain (TFIC) is a paradigmatic, exactly solvable example of a 1-D quantum phase transition (QPT)~\cite{qpt_Sachdev} and has garnered significant recent interest in the study of entanglement and scrambling dynamics~\cite{Lin2018}, dynamical phase transitions ~\cite{DPT_Gorshkov_2019}, and quantum thermalization~\cite{Calabrese_2011}. This model has been experimentally realized using Rydberg atoms when interactions beyond nearest neighbors can be neglected on the relevant timescales~\cite{Labuhn_2016, Bernien_2017, Guardado_2018}. The Hamiltonian for the model is

\begin{equation}
    H_{TFIC} = -J \sum_{i=1}^L \sigma^z_i\sigma^z_{i+1} + h \sum_{i=1} \sigma^x_i.
    \label{Ising}
\end{equation}
where periodic boundary conditions have been employed such that $\sigma^\mu_{L+1} = \sigma^\mu_1$. This model exhibits a ground state phase transition at $g=h/J = 1$ that is in the same universality class as the classical 2-D Ising model. The critical point $g_c$ separates two distinct phases, where for $g<g_c$, the spins exhibit $\mathcal{Z}2$ symmetry-breaking order with nonzero longitudinal magnetization $\langle\sigma^z\rangle$. For $h>h_c$ the spins are in a paramagnetic phase with preferential alignment along the transverse field. The system also exhibits a dynamical-QPT (DQPT) when the system is prepared in an initial product state $|\uparrow \uparrow \uparrow...\rangle_z$ and quenched across the ground state critical field~\cite{DPT_Gorshkov_2019}.


In our work we employ an ancillary, central qudit to probe the TFIC. Depending on the underlying phase and transport characteristics, novel real-time dynamics and long-time behavior arise in the auxiliary system. 
The form of the spin-ancilla interaction is a paradigmatic Dicke-Ising construction that combines an essential model for light-matter interactions and magnetic quantum matter represented by the following:

\begin{equation}
    H_{int} = \omega_c a^\dagger a + \frac{\lambda}{\sqrt{L}}\sum_{i=1}^L (a^\dagger + a )\sigma^x_i,
    \label{Dicke}
\end{equation}
where $a (a^\dagger)$ is a bosonic annihilation (creation) operator for a single mode uniformly coupled to each spin with strength $\lambda$ and normalized by $\sqrt{L}$ such that the effective spin-spin interaction induced by the bosonic ancilla remains intensive. The second term in Eq.\ref{Dicke} represents the rotation of a single spin in exchange for the particle number within the ancilla. Though here we specifically consider bosonic fields with finite dimension $q > L$, the problem is identical to a large spin of size $S = q/2$ with Zeeman splitting $\omega_c$.

We first address the general behavior of spin-ancilla observables and how they deviate from both the pure Ising and the Dicke models. The general relationship between observables is obtained from the exactly solvable ground state in the $(J=0)$ limit where the bosonic occupation $\hat{N}$ is 
\begin{equation}
    \langle \hat{N} \rangle \propto \frac{\langle \hat{S}_x^2\rangle \lambda^2}{\omega_c},
    \label{scaling}
\end{equation}
where here $\hat{S}_x = \sum_i \sigma^x_i$. This can be readily observed by performing a unitary transformation on Hamiltonian by translating the bosonic creation and annihilation operators as
\begin{equation}
    \hat{b} = \hat{a} + \frac{\lambda}{\sqrt{L}\omega_c}\hat{S}_x.
\end{equation}
The resulting Hamiltonian has no spin-boson coupling and can be treated as a classical Ising chain
\begin{equation}
    \hat{H} = \omega_c \hat{b}^\dagger \hat{b} + \frac{\lambda^2}{L\omega_c}\hat{S}_x^2 + h\hat{S}_x.
    \label{eq:exactSB}
\end{equation}The relevant timescale translating the exchange of spin and bosons is the ancilla splitting $\omega_c$, which for all results in this work is set to $2\pi J/ \omega_c = 12.6tJ$. The system conserves total $S_x$ so when prepared in an eigenstate of $S_x$: $\psi = |m\rangle \otimes |0\rangle$, where $|0\rangle$ represents bosonic vacuum, the spin system will not undergo dynamics following a quench, while the bosonic ancilla will fluctuate between $|0\rangle$ and $|\alpha_m\rangle$ at the characteristic level splitting frequency. $|\alpha_m\rangle$ is a coherent state centered at $m^2 \frac{\lambda^2}{\omega_c}$. Once Ising interactions ($|J|>0$) are included along with the nonlocal central ancilla coupling, the model is no-longer exactly solvable, but we anticipate in the low energy regime that same characteristic behavior captures the Ising-boson ground state and low energy quenches. The spin-ancilla interaction will shift the underlying ground state and dynamical quantum phase according to the effective mean-field magnetic interaction $\frac{\lambda^2}{\omega_c}$. Greater details on the physics of the Ising-Dicke model and magnetic phase transition in the absence of a transverse field can be found in a recent work by Rohn et al.~\cite{rohn2020}.


\begin{figure}
    \centering
    \includegraphics[width=0.45\textwidth]{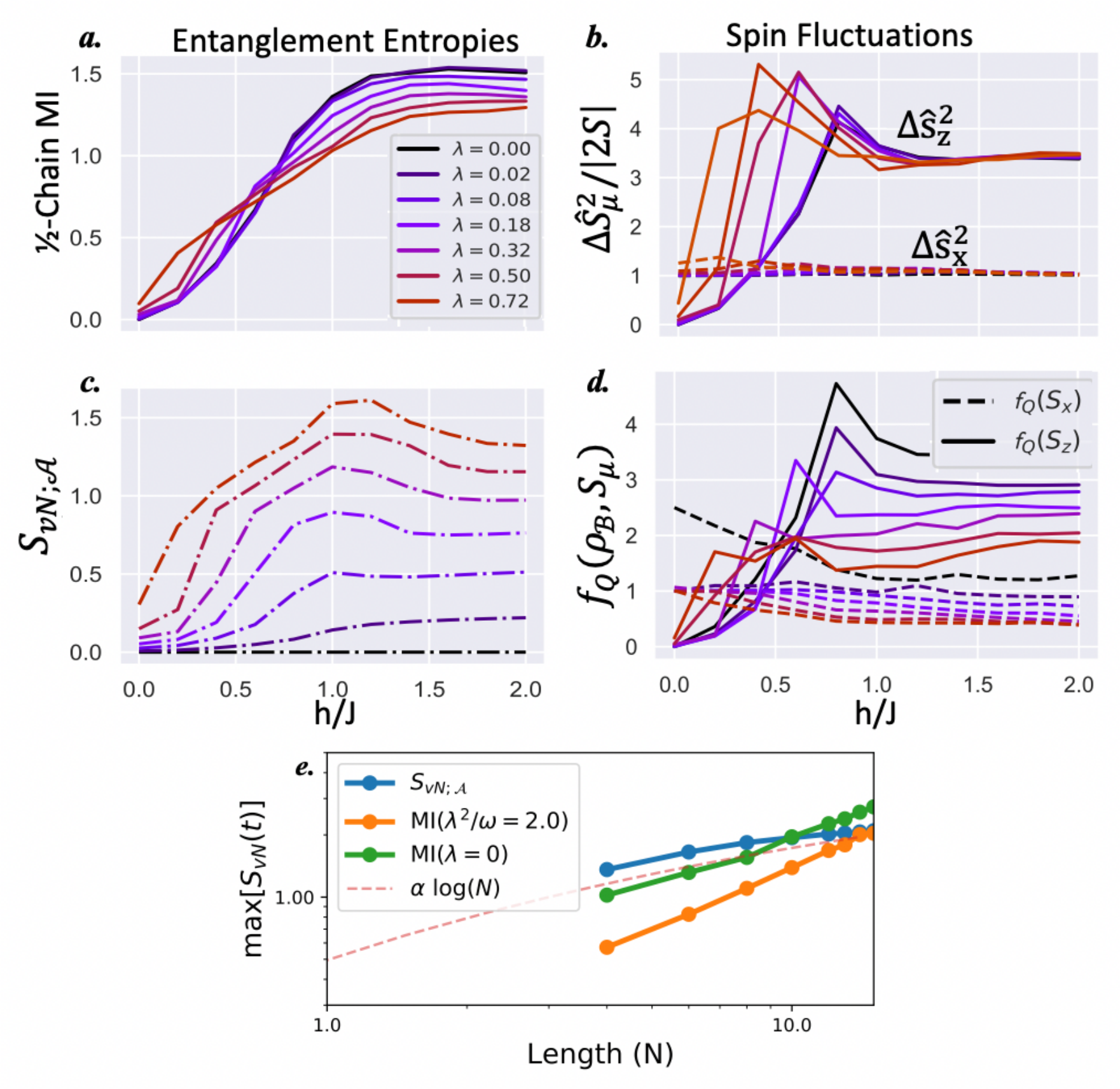}
    \caption{{\bf Entanglement metrics:} Long-time average entanglement: half-chain MI (a), spin-ancilla von Neumann entanglement entropy $S_{vN, \mathcal{A}}$ (b), spin fluctuations $\Delta s_{\{x,z\}}^2$ (c), and QFI along $S_{\{x,z\}}$ (d). Results presented as a function of transverse magnetic field $h/J$ and $\lambda$ following a quench from unentangled spin-boson product state. (a) MI grows like the number of excitations present in the initial quench and saturates at high field at a value that scales with the system size. 
    Entanglement saturation value decreases with increasing $\lambda$ and profile shifts to lower field $h$ as ancilla modifies DQPT. (b) Ancilla $S_{vN,\mathcal{A}}$ picks up the signature of the dynamical phase transition with a peak near the finite-size resolved critical point and a saturation entropy that grows proportionally to $\lambda$ in the paramagnetic phase. (c) $\Delta s_z^2$ grows similar to the MI; initially zero in the product state and saturates to a constant density $\sim 3$. $\Delta s_z^2$ has a finite-size resolved peak about the critical point, modified with growing $\lambda$. $\Delta s_x^2$ remains constant as a function of $h$ and $\lambda$. Spin fluctuations provide an upper bound (Eq.\ref{eq:pQFI}) for the respective QFI measures (d). Longitudinal and transverse $f_Q$ monotonically decreases as a function of $\lambda$. (e) Finite size scaling in the paramagnetic phase $h = 2.0$ of $S_{vN, \mathcal{A}} (q = 20, \lambda^2/\omega_c = 2.0$) and MI in the zero-coupling and strong coupling regime $\lambda^2/\omega_c = 2.0$. MI shows characteristic volume-law entanglement scaling even in the presence of the external environment while $S_{vN, \mathcal{A}}$ shows slow growth with system size and is accurately described by $S_{vN, \mathcal{A}} \propto \log(L)$. System size $L=12, 8$ for (a-c, d), finite ancilla dimension $q=40$, $J=-1, \omega_c = 0.5$, and $tJ\in[0,50]$.}
    \label{fig:quench_ent_cavity}
\end{figure}

\begin{figure*}[htb!]
    \centering
    \includegraphics[width=0.94\textwidth]{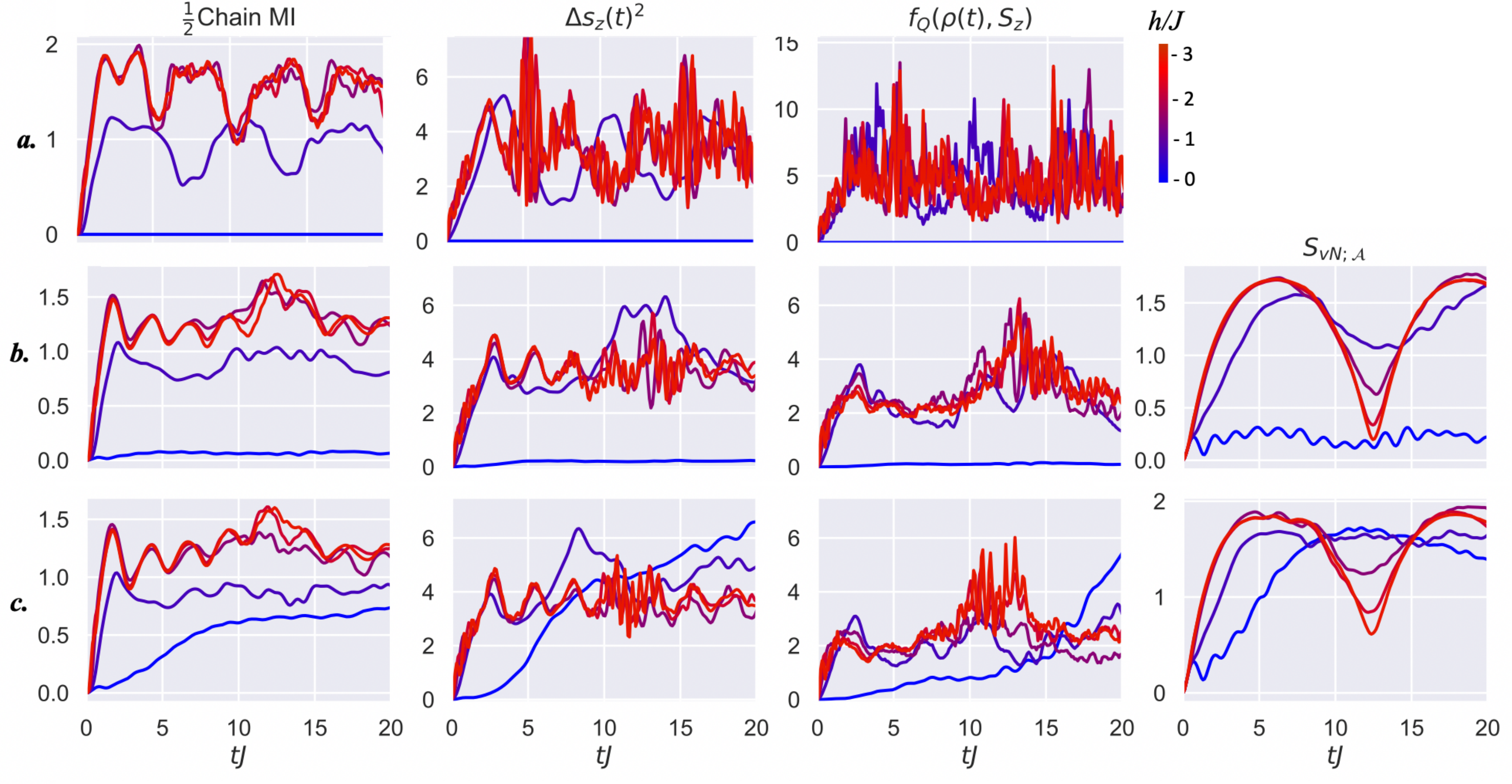}
    \caption{{\bf {Real-time Entanglement Dynamics:}} Entanglement metrics following a quench from the polarized state: $MI(L/2)$, $\Delta s_z(t)^2$, $f_Q(S_z)$, and $S_{vN ; \mathcal{A}}$. The three rows ((a,b,c) correspond to $\lambda^2/\omega_c = \{0, 0.63, 1.13\}$). (a) For $\lambda = 0$, MI, spin-fluctuations and QFI grow with time $t \sim L/v_{max} = 2.5/J$, after which both oscillate about an average value that grows with field and saturates above the DQPT critical point. As $\lambda$ increases, $S_{vN; \mathcal{A}}$ grows in magnitude and similarly oscillates like the characteristic period given by the bosonic level spacing $12.6tJ$. The maximum ancilla entanglement similarly reaches a saturation threshold across the DQPT. Increasing $\lambda$ increases the initial MI and spin-fluctuation growth rate but decreases saturation value for quenches into the polarized phase. The oscillations across all spin density matrix measures (most dramatic across the DQPT in red) are most rapid for times $t\approx \tau = 12.6tJ$, the ancilla period. 
    The feature of greatest interest in the growing difference between $\Delta s_z(t)^2$ and $f_Q(S_z)$. In (a) we see that, up to numerical noise in measuring the diagonal operator $S^z$ over the spin reduced density matrix, the profiles are identical as expected. Moving to (b) and (c) the difference grows with $\lambda$ regardless of spin phase. System size $L=10, d = 30$ and $h$ values $(0, 0.75, 1.5, 2.25, 3)$.}
    \label{fig:nequench_ent_rt}
\end{figure*}

\medskip

\medskip

\section{Entanglement Dynamics}
\subsection{Long-time Average Entanglement}
The spin-ancilla coupling makes the Ising Hamiltonian nonintegrable, and in studying its dynamics we employ exact diagonalization (ED) and a real-time evaluation of the Schrodinger ODE. We study how a fully polarized, non-equilibrium product state behaves when quenched by Eqs.\ref{Ising},\ref{Dicke}. This study sheds light on how the spin-chain QPT evolves under a highly non-local coupling, how quantum fluctuations in many-body systems lead to eventual equilibration, and how composite information dynamics develop as a result of this 2-body (chain-ancilla) construction. We study the real-time dynamics of the full spin-ancilla density matrix and measure spin-ancilla entanglement, spin-spin mutual information, and the QFI contained in the spin reduced density matrix.

We initially prepare the Ising spin-chain and the ancilla bosonic system in an unentangled product state
$
    |\psi(t=0)\rangle = |\psi_{\text{s}}\rangle \otimes |0\rangle
$ 
with $|0\rangle$ representing bosonic vacuum and an unentangled polarized state $|\psi_{\text{s}}\rangle = |\uparrow\uparrow\uparrow...\rangle$. 
A quench is then performed at $t=0$ from the initial wavefunction to $(J=-1, h/J, \lambda)$ where we then vary $h$ and $\lambda$.

\medskip

In the non-equilibrium quench scenario first ignoring the ancilla $\lambda = 0$, the half-chain MI transitions from $0$ at $h=0$ toward a volume law entangled state across a critical magnetic field (Fig.\ref{fig:quench_ent_cavity}a.). Deep into the polarized phase, the initial state's energy begins to lie within the center of the spectrum, essentially becoming a highly excited pure state (Generalized Gibbs Ensemble GGE) with extensively scaling entropy. Excitations encoded into the initial non-equilibrium state propagate and distribute entanglement~\cite{Calabrese2005}. The saturation in entanglement occurs concurrently with the saturation in excitations as exhibited by the steady-state domain wall count~\cite{Tan2021}. The entanglement saturation value is an interesting focus of study, where the scaling is indicative of the integrable or nonintegrable character of the model and additionally a signature of many-body localized systems~\cite{Bardarson2012}.

Spin-fluctuations and QFI (identical in the $\lambda = 0$ limit) seen in Figs.\ref{fig:quench_ent_cavity}(b,d) depict the same entanglement growth; both saturating for quenches across $h/J > 1$. Though entropy and multipartite entanglement both depict extensively scaling entanglement in the saturated regime, the entanglement entropy has a smooth crossover, in agreement with previous results for larger systems, while QFI depicts a second order transition and peak about the ground state critical point $h = J$. The $\Delta\hat{S}_z^2$ results in Fig.\ref{fig:quench_ent_cavity}b. agree with previous theoretical results in the paramagnetic regime that show $\Delta\hat{S}_z^2(h/J>1, t = \infty) = 3$~\cite{Pappalardi_2017}. Further discrepancies arise due to the difference in polarized vs. cat-state initial spin-state. 

When additionally quenching the system with nonzero $\lambda$, identical entanglement behavior is imprinted on the ancilla entanglement entropy (Fig.\ref{fig:quench_ent_cavity}c.), where long-time entanglement grows toward the critical field and saturates across the DQPT. The ancilla shifts the critical field and at moderate coupling $\lambda^2/\omega_c$ decreases the relative entanglement surrounding the critical point, observed in Fig.\ref{fig:quench_ent_cavity}a. MI increases slightly in the ferromagnetic phase as the underlying ground state shifts to lower magnetic fields. In the saturated phase $h/J>1$, MI and QFI (Fig.\ref{fig:quench_ent_cavity}a. and d.) are exclusively removed from the spin subsystem, which leads to a monotonic decrease in the saturation value with coupling. The decrease in spin-chain entanglement occurs concurrently with growing $S_{vN}$ between spins and ancilla. The most interesting observation is that this loss is not observed in the spin fluctuations (Fig.\ref{fig:quench_ent_cavity}c.), where the saturation value maintains a fixed value $\sim 3$. It is intuitive to think that saturated entangled DoF in the paramagnetic quench share increasing information with the ancilla with growing $\lambda$, removing previously saturated spin-spin entanglement and redistributing excitations between spin and ancilla. This is the same behavior observed in the spin-fluctuations and domain-wall density profiles, which give credence to a possible relationship between excitations and environmental entanglement. The most interesting discovery is that MI continues to scale with system size while spin-ancilla entanglement grows at a much slower rate $\sim \log(L)$ regardless of coupling (Fig.\ref{fig:quench_ent_cavity}e.).

This picture persists even when varying the qudit size [see Supplemental for further details]. As the Hilbert space $d$ decreases, the maximum amount of information entropy that can be stored between the spin and ancilla decreases as $\log(d)$, so it is expected that a smaller ancilla will have a similarly decreasing impact on entanglement within the spin-subsystem. This entanglement loss picture raises important questions: how can we numerically evaluate the spin-ancilla entanglement based off of measurements on the spin system? How do different phases limit or share stored information? How does this description hold up in real-time?

\medskip

\subsection{Real-time Numerical Results}
In the non-equilibrium quench, the initial state has trivial entanglement characteristics and reaches a maximum on the order of the maximum quasiparticle velocity $L/v_{max}$ where $L=10$ gives $t=2.5J /\min{[1,h]}$. When $\lambda = 0$ [Fig.\ref{fig:nequench_ent_rt}(a)] we see initial growth followed by oscillations connected to the integrability of the model and finite size effects. As $\lambda$ increases [Fig.\ref{fig:nequench_ent_rt}(b,c)], oscillations in MI, spin-fluctuations, and QFI decrease on the order of the spin-ancilla entanglement entropy profile. More interestingly, we observe a deviation in the QFI profiles from the spin fluctuations, as compared to the exaggerated behavior shown in Fig.\ref{fig:cQudit}(c). MI and QFI in the paramagnetic quench decay from the $\lambda = 0$ results with revivals on the order of the cavity frequency $\omega_c$. $\Delta S_z(t)^2$ on the other hand maintains a strong average value for all times, taking into account the shift in underlying critical field. In contrast to the decaying QFI profile, ancilla entanglement initially grows rapidly and subsequently decays, oscillating at its characteristic timescale for quenches in the paramagnetic regime.

Average spin fluctuations remain largely unchanged as a function of $\lambda$ in the real-time evolution as well as at long-times, while QFI shows a clear modification in the non-interacting to interacting ancilla limit. The developing discrepancy between fluctuations and QFI in the presence of environmental entropy suggest a good fixed point to understand how information is distributed between spins and the ancilla and how $S_{vN; \mathcal{A}}$ provides greater insight in the the inequality in Eq.\ref{eq:pQFI}.

\medskip

\begin{figure}
    \centering
    \includegraphics[width=0.5\textwidth]{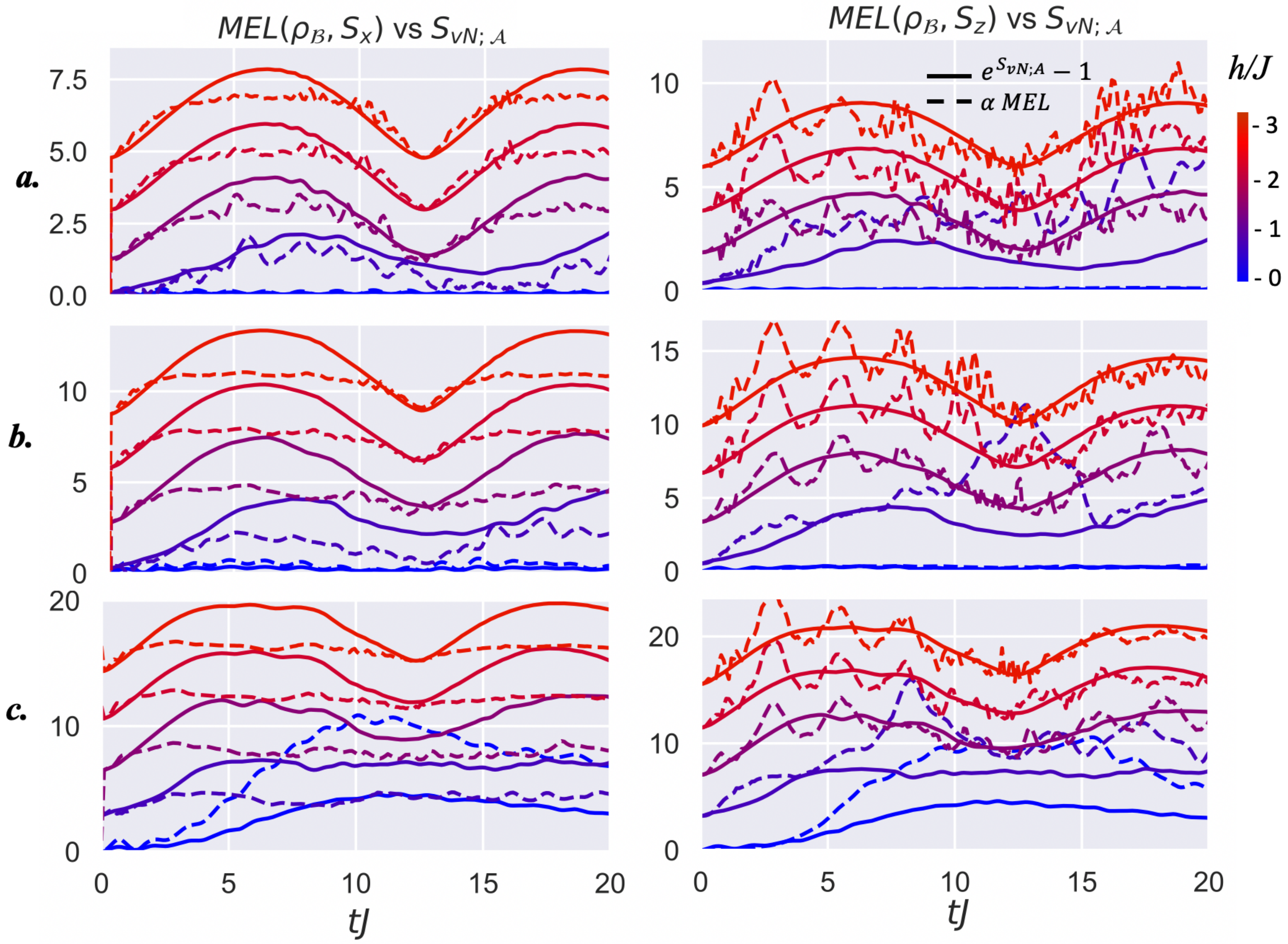}
    \caption{{\bf {Multipartite Entanglement Loss:}} Real-time evolution of the transverse and longitudinal (left, right)  multipartite entanglement loss (MEL) (dotted) vs ancilla $S_{vN}$ (solid), $\lambda^2/\omega_c = 0.28, 0.63, 1.13$ (a,b,c) from polarized initial state $|\uparrow\rangle^{\otimes N}|0\rangle$. Curves offset vertically for clarity. Fisher information is reduced compared to the upper bound set by the fluctuations along both spin vectors, leading to the growth in $MEL$ and ancilla entanglement entropy. $MEL(S_x)$ and $MEL(S_z)$ provide extremely accurate estimates of the ancilla entanglement entropy for $\lambda^2/\omega_c << J$ regardless of underlying phase. With increasing coupling, the amount of information gained by the ancilla surpasses the multiparticle entanglement as measured by $f_Q(S_x)$, saturating $MEL(S_x)$. As more entanglement is encoded in $\hat{S}_z$, $MEL(S_z)$ continues to provide a strong estimate on the ancilla entanglement and is nearly exact in the paramagnetic, saturated phase $h>J$. System size $L=10, d=30$ and $h$ values $(0, 0.75, 1.5, 2.25, 3)$.}
    \label{fig:neq_MEL_M}
\end{figure}

\subsection{Quantifying Entanglement Loss}
In the quench scenario, Ising spins entangle rapidly in time as observed by the development of MI and QFI. As coupling with the ancilla increases we see a deviation from fluctuations and multipartite entanglement like the ancilla entanglement entropy profile. Where fluctuations remain consistent (Fig.\ref{fig:quench_ent_cavity}(b)), spin-spin QFI and ancilla entanglement undergo highly nontrivial behavior as a function of the system-environment entropy. We define multipartite entanglement loss $MEL(\rho_\mathcal{B}, \hat{O})$ to capture the difference in observable fluctuations from the true multiparticle spin entanglement in the reduced density matrix over the spin-chain $\rho_\mathcal{B}$. Here we calculate $MEL$ along the longitudinal and transverse collective spin vector $S_\mu = \sum_i \sigma^\mu_i; \mu \in [x,z]$. These quantities were previously explored using exact solutions to the Ising model and, though may not represent $S_{max}$ as we have suggested in Eq.\ref{eq:result}, provide two QFI measures that accurately capture the dynamical Ising phases~\cite{Pappalardi_2017}. 

For weak ancilla couplings [Fig.\ref{fig:neq_MEL_M}(a)] we see that both $MEL(S_x)$ and $MEL(S_z)$ serve to capture the real-time behavior of the ancilla entanglement as well as accurately capture the amplitude of information gain. As coupling increases and the ancilla begins to sink more entanglement from the Ising spins, $S_{vN;\mathcal{A}}$ begins to saturate the entanglement loss observed through $MEL(S_x)$ Fig.\ref{fig:neq_MEL_M}(b,c: left). $MEL(S_x) < e^{S_{vN; \mathcal{A}}} - 1$ is most noticeable at half of the ancilla level splitting $\frac{\pi}{\omega_c} = 6.28tJ$. Though $MEL(S_x)$ serves as a poor estimate of spin-ancilla entanglement at large couplings and for $h>J$, $S_{vN; \mathcal{A}}$, $MEL(S_z)$ profiles remain in strong agreement especially in the paramagnetic phase. The transition from $MEL(S_x) \sim MEL(S_z) \sim e^{S_{vN; \mathcal{A}}} - 1$, to  $MEL(S_x) < MEL(S_z) \sim e^{S_{vN; \mathcal{A}}} - 1$ can be attributed to the larger spin fluctuations in $S_z$. In the paramagnetic phase $\Delta S^2_z = 3L$ and $\Delta S^2_x = L$, so as ancilla entanglement grows, the loss perceived by $MEL(S_x)$ is artificially low since less entanglement is observed along this direction. As $\Delta S^2_z$ is the direction of maximal fluctuations/QFI, we anticipate that $MEL(S_z)$ remains a strong ancilla entropy estimate. This is indeed the case where in Fig.\ref{fig:neq_MEL_M}(b,c: right) entropy remarkably traces $MEL(S_z)$ even beyond $\lambda^2/\omega_c > J$. 

In all of our results $S_{vN; \mathcal{A}}$ is never parametrically larger than $\max[MEL(S_x, t), MEL(S_z, t)]$ with the greatest discrepancy developing about the critical point. Surrounding the critical point we expect the transverse and longitudinal QFI to present an inaccurate characterization of information loss as maximal spin-variation generally lies in the $x-z$ plane. On the other-hand, in the large-field limit, we expect this nonintegrable system to reach a relatively stable steady-state as a function of field and ancilla coupling, so $MEL(S_z)$ should continue to paint an accurate picture. Deep in paramagnetic regime, where the system reaches an approximate infinite temperature state and approaches proximate $S_x$ conservation ($J < h < \lambda^2/\omega_c$), QFI along $S_z$ will continue to dominate.

\section{Discussion}

Here we have extended the scope of how information sharing of entangled degrees of freedom leads to a nearly direct connection between spin fluctuations, Fisher information, and external entanglement entropy. Our characterization of MEL and the numerically observed proportionality with $S_{vN;\mathcal{A}}$ serves as an extension of the exact results observed in quadratic bosonic/fermionic systems~\cite{Hackl2018, Bianchi2018} and behavior seen in semiclassical, infinite/long-range spin systems~\cite{lerose2020}. These results also agree with numerical sampling results over small Hermitian matrices~\cite{toth2018lower}, and, here, due to the collective spin-boson interaction we connect this collection of works by finding $\text{MEL} \propto S_{vN;\mathcal{A}} \propto \log{L}$. We find that spin subsystem-fluctuation measurements provide a strong estimate on establishing the number of entangled DoF even in composite systems. In Fig.\ref{fig:neq_MEL_M} we show that entanglement can be stored directly in local spin fluctuations when $MEL = 0$ or it can be stored in shared degrees with an environment $MEL \sim e^{S_{vN; \mathcal{A}}} - 1$. This novel characterization not only presents a qualitative framework that captures the long-time equilibration dynamics in our composite spin-ancilla system but correlates directly with the real-time entropy fluctuations between spins and ancilla. We observe that when the ancilla acts purely as a bath and $\Delta S^2_\mu (t, \lambda) \leq \Delta S_\mu^2 (t, 0)$, MEL perfectly captures entanglement physics between spins and ancilla. In the highly energetic regime where the Ising chain is fully excited, the ancilla can only serve as an effective bath, whereas in the low-energy quenches for $h << J$, the system can be highly excited by ancilla interaction. In the low-field quench, MEL tends to be parametrically larger than entanglement which suggests a possible upper bound in Eq.\ref{eq:result} may be more accurate.


Our results are in good agreement with recent works that establish a similar picture of spin-fluctuations and entanglement in semiclassical models. In the Dicke model and its corresponding regular to ergodic phase transition it was shown that spin-boson entanglement similarly is responsible for the development of additional collective spin-fluctuations, and in the kicked Ising system collective squeezing occurs in tandem with entanglement entropy between bipartitions~\cite{Gietka_2019, lerose2020}. Though previous research presents an intuitive picture, there is no established/quantitative differentiation between the internal multipartite spin entanglement and the composite system-environment entanglement contribution in the development of physical correlations. Our results provide a deeper heuristic examination into how fluctuations and QFI relate in a simple, composite model.

This novel characterization of entanglement physics in a composite spin-boson model provides an intuitive description on how DoF are shared with different subsystems in real-time and helps provide a generalized picture on information. We see that the proliferation of excitations that encode collective fluctuation play a critical role in the strength of entanglement entropy response when probed by an external, homogeneous ancilla. The collective interaction relays only semiclassical information $S_{vN;\mathcal{A}} \propto \log{L}$ regarding the dynamics of the local, interacting quantum spin system and presents a novel translation of information. Though this purely numerical investigation focuses on highly excited states of the Ising model and a particular transverse coupling, this physics remains consistent in low energy quench scenarios about the Ising ground state as a function of $h/J$ [see Supplemental for greater details].

\section{Conclusion}

From our results, experimental work capable of measuring global spin-fluctuations and cavity/bosonic correlations can develop a richer characterization of the full information dynamics of multiparticle, composite systems. Where determining QFI in an arbitrary, non-equilibrium, mixed system requires full state tomography, our work shows that MEL provides a simple extension for pure states by considering external and internal subsystem entanglement content. Using this method, we recreate effective estimates for the amount of entangled information contained in any subsystem of a pure quantum ensemble. We anticipate our results to have an impact on experimental quantum information studies in cavity-QED and NV-center platforms, where the range of interaction consists of a macroscopic region of the system of interest.

This work focuses on adding an environment to the well-understood Ising model, where quantum thermalization, entanglement, and spin fluctuations have been thoroughly characterized. It would be an interesting future pursuit to explore less well understood models, or fully nonintegrable models where the optimal QFI measure is not known, and observing the entanglement consequences in a similar system-environment construction. Exploring semiclassical models where analytic forms of QFI, MEL, and $S_{vN}$ can be explored would provide a wealth of insight into the physics observed in this local quantum system. It is also necessary to investigate how this semiclassical transfer of information breaks down with either inhomogenous bath couplings or a larger/more complex environment and leads to full thermalization. With regards to thermalization, we have interestingly seen that excitations and QFI are intimately related as dynamical quasiparticles encode fluctuations, but at the same time they seem to dictate the entanglement response to an external environment. It would be interesting to develop a more generalized fluctuation dissipation relations regarding entanglement susceptibility and excitation exchange/production. As we have seen in our results, MI remains largely preserved in the spin-system and only decreases by $\sim\log{L}$ compared to the noninteracting case. A similar phenomenon was observed in an open Ising chain where the system did not completely thermalize $MI(L/2, t\rightarrow \infty) > 0$~\cite{maity2020}; so how does the form of the system-environment interaction preserve certain quasiparticles/operators propagation and encode entanglement entropy within the system.

Our work additionally brings up interesting metrics for determining how to engineer and transfer quantum information between subsystems; where we initially store or grow multiparticle entanglement from the ground state or non-equilibrium quench and then transform it into external entanglement entropy. Future work will explore the genuine multiparticle entanglement mediated by an ancilla and how it is bounded by the size of the ancilla Hilbert space. A parallel investigation as to how information is scrambled with a weak non-local central vertex shows promise in understanding fast scrambling dynamics, where no work has been done on relating operator growth rate and the size of the ancilla dimension~\cite{szabo_unpublished}. In the large cavity limit, cavity interactions generate all-to-all spin interactions and lead to tunable OTOC physics in the spin subsystem~\cite{marino2019}, but no work has examined the possible fingerprint imprinted on the central mode. This would provide fruitful scrambling protocols and analyses amenable to cQED and NV center experiments. Finally it would be interesting to see if topological signatures inherent in quantum fluctuations or full system entanglement content leave unique signatures in the shared information with an external observer. Recent work has focused on the entanglement contribution of quasiparticles above a ground state~\cite{wybo2020, you2020} and our future work will explore the connections between symmetry preserving excitations, fluctuations, and entanglement dynamics.

\section{Acknowledgements}
We would like to thank Elan Shatoff, Bradley Goff, Franz Utermohlen for feedback and review of the manuscript as well as Sayantan Roy and Zachariah Addison for useful discussions. This material is based upon work supported by the U.S. Department of Energy, Office of Science, Office of Basic Energy Sciences under Award Number DE-FG02-07ER46423. Computations were done on the Unity cluster at the Ohio State University.

\bibliography{references}


\end{document}


\title{Supplemental Material for "Entanglement Dynamics between Ising Spins and a Central Ancilla"}
\author{Joseph C. Szabo}
\affiliation{Department of Physics, The Ohio State University, Columbus, Ohio 43210, USA}

\author{Nandini Trivedi}
\affiliation{Department of Physics, The Ohio State University, Columbus, Ohio 43210, USA}

\maketitle

\newpage
\onecolumngrid
\setcounter{section}{0}
\renewcommand{\thefigure}{S\arabic{figure}}
\setcounter{figure}{0}

\section{TFIC Exact Solution}
The TFIC is exactly solvable following a Jordan-Wigner mapping to non-interacting spins and subsequent Fourier and Bogoliubov transformation. \begin{align}
    H &= -J (\sum_{\langle ij \rangle}\sigma^z_i\sigma^z_j - h\sum_i{\sigma^x_i}) \\
    \sigma^x_i &= 1 - 2c^\dagger_ic_i \\
    \sigma^z_i &= -\prod_{j<i}\sigma^x_j (c^\dagger_i + c_i) \\
    H^{\small{JW}} &= J (\sum_{\langle ij \rangle}c^\dagger_ic_j + c_ic_j + H.C. + h\sum_i 1 - 2c^\dagger_ic_i)
\end{align}

Then Fourier transforming the fermionic operators gives us:

\begin{align*}
    d^\dagger_k &= \frac{1}{\sqrt{L}}\sum^L_j e^{i(2\pi jk)/L} c^\dagger_j, \\
    d_k &= \frac{1}{\sqrt{L}}\sum^L_j e^{-i(2\pi jk)/L} c_j ,\\
    H^{\small{JW}} &= J \sum_k \large{(}[h - \cos{\frac{2 \pi k}{L}}]d^\dagger_k d_k \\
    & - \frac{i}{2}\sin{\frac{2 \pi k}{L}[d_{-k}d_k + d^\dagger_{-k} d^\dagger_k] - h/2}\large{)}, \\
    b^\dagger_k &= u_kd^\dagger_k + iv_kd_{-k}, b_k = u_kd_k + iv_kd^\dagger_{-k}, \text{ where}\\
    u_k &= \cos{\theta_k}; v_k = \sin{\theta_k}; \tan{\theta_k} = \frac{J \sin{k}}{B - J \sin{k}} \\
    H^{BdG} &= \sum_k \epsilon_k b^\dagger_k b_k + \sum_k \epsilon_k 
\end{align*}

This gives us the exact spectrum for the transverse field Ising model with ground state energy that can be characterized by integrating over the dispersion relation given as

\begin{equation}
    \epsilon_k = J \sqrt{1 + h^2 - h\cos{k}}.
\end{equation}

The ground state of the model can be determined from the empty fermionic vacuum state $|0\rangle \sim |\rightarrow\rightarrow\rightarrow\rightarrow...\rangle$ (in terms of the spins) and eliminating all momenta from the vacuum state: $|\Psi_0 \rangle = \Pi_k b_kb_{-k}|0\rangle$. 

The initial state ($t<0$) for a sudden quench experiment will occupy excited modes of the Hamiltonian for $t>0$. For a sudden quench of the Hamiltonian we can consider that $|\Psi_0\rangle$ is the ground state of $H_0$. For systems with an extensive number of conserved quantities i.e. integrable systems, the infinite time steady state thermalizes according to a Generalized Gibbs Ensemble~\cite{Calabrese_2011}. This state can be decomposed into the rapidities or conserved quantities (occupation of the eigenmodes) of the time-evolution Hamiltonian, written as

\begin{align}
    |\Psi_0\rangle &= \sum_k c(k) |n_k\rangle, \\
    c(k) &= \langle \Psi_0 | n_k | \Psi_0 \rangle, \\
    |\Psi(t)\rangle &= e^{iHt}|\Psi_0\rangle, \\
    |\Psi(t)\rangle &= \sum_k e^{i\epsilon_kt}c(k)|n_k\rangle
\end{align}

For all occupied excited modes $n_k$ we can think of an excitation with energy $\epsilon_k$ being present within the spin chain. With the excitation present at $t = 0$, this will propagate at a finite velocity. The velocity of this excitation is given by the dispersion relation taken from the spectrum. This is identically what determines the velocity of entanglement growth in the integrable Ising model~\cite{Alba2017, bastianello2018}.

\begin{align}
    v &= \frac{d\epsilon_k}{dk} \\
    v_{\text{max}} &= \text{max}[\frac{Jh\sin{k}}{\sqrt{1 + h^2 - 2h\cos{k}}}]\\
    &= J \ \text{min}[1, h]
\end{align}

\begin{figure}
    \centering
    \includegraphics[width=0.345\textwidth]{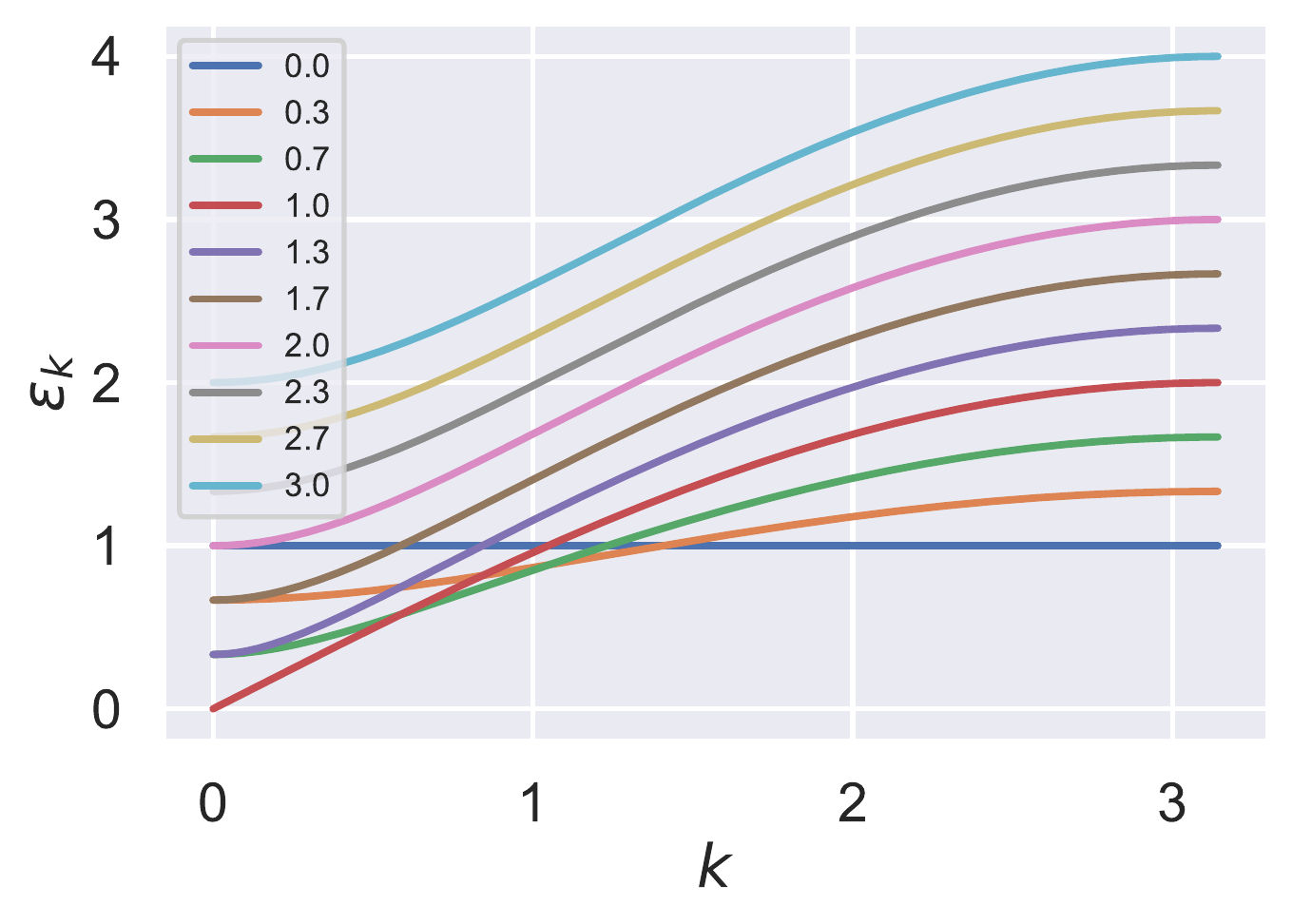}
    \includegraphics[width=0.345\textwidth]{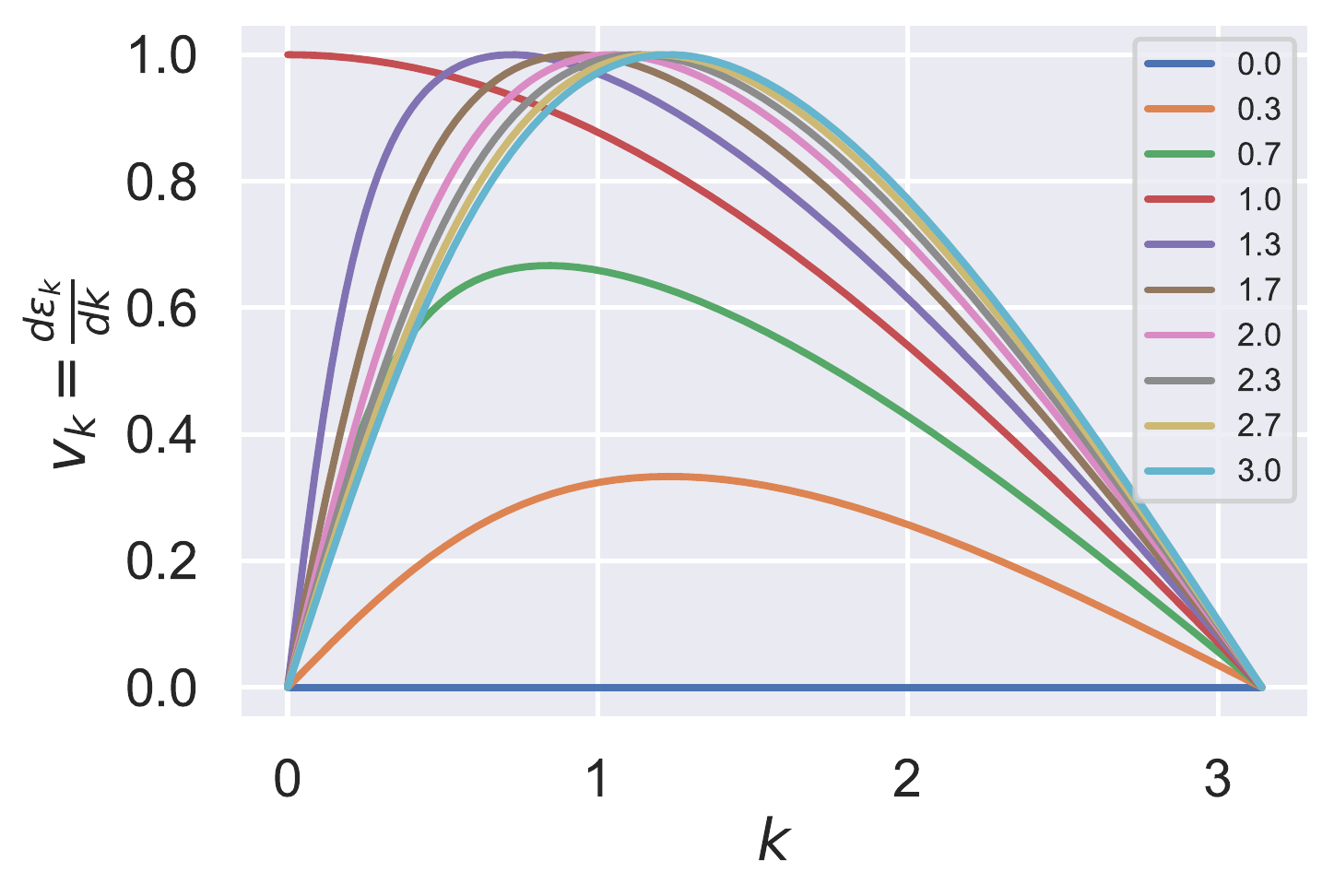}
    \caption{
    TFIM energy spectrum (left) and group velocity for excitations (right) of eigenmode $k$ as a function of transverse magnetic field $h/J$.}
    \label{fig:ising_spec}
\end{figure}

\section{Finite Size and Additional Entanglement Loss Results}

\begin{figure}
    \centering
    \includegraphics[width=0.95\textwidth]{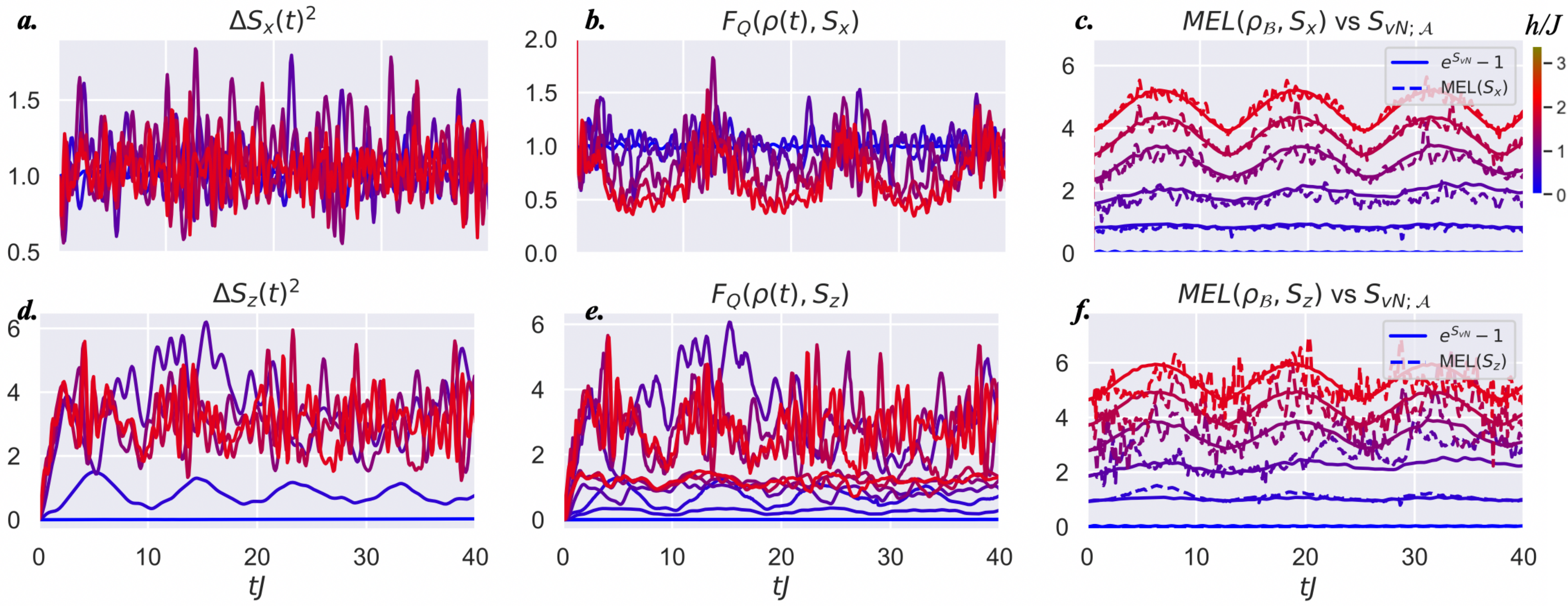}
    \caption{\textbf{MEL dynamics from non-equilibrium quench $(\lambda^2/J\omega_c = 0.125)$}. Real-time evolution of the transverse and longitudinal (top, bottom) spin-fluctuations (a,d), $\mathcal{f}_Q$ (b,e), and multipartite entanglement loss (MEL) (dotted) vs ancilla $S_{vN}$ (solid) (c,f) from polarized initial state. Rapid spin fluctuations are suppressed on the order of ancilla entanglement profile in both $x,z$ spin components. Fisher information is reduced compared to the upper bound set by the $\Delta s_\mu^2(t)$ along the two spin vectors. (c,f) Taking the difference between fluctuations and QFI provides a good estimate on the ancilla entanglement entropy. For weak coupling and numerical noise associated with $f_Q(S_z)$, $MEL(S_x)$ provides a good estimate for all spin phases. MEL and $S_{vN}$ curves are offset vertically for clarity. System size $L = 8, d = 40$ and $h$ values $(0, 0.4, 0.8, 1.2, 1.6, 2.0)$.}
    \label{fig:neq_MEL_W}
\end{figure}

\begin{figure}
    \centering
    \includegraphics[width=0.95\textwidth]{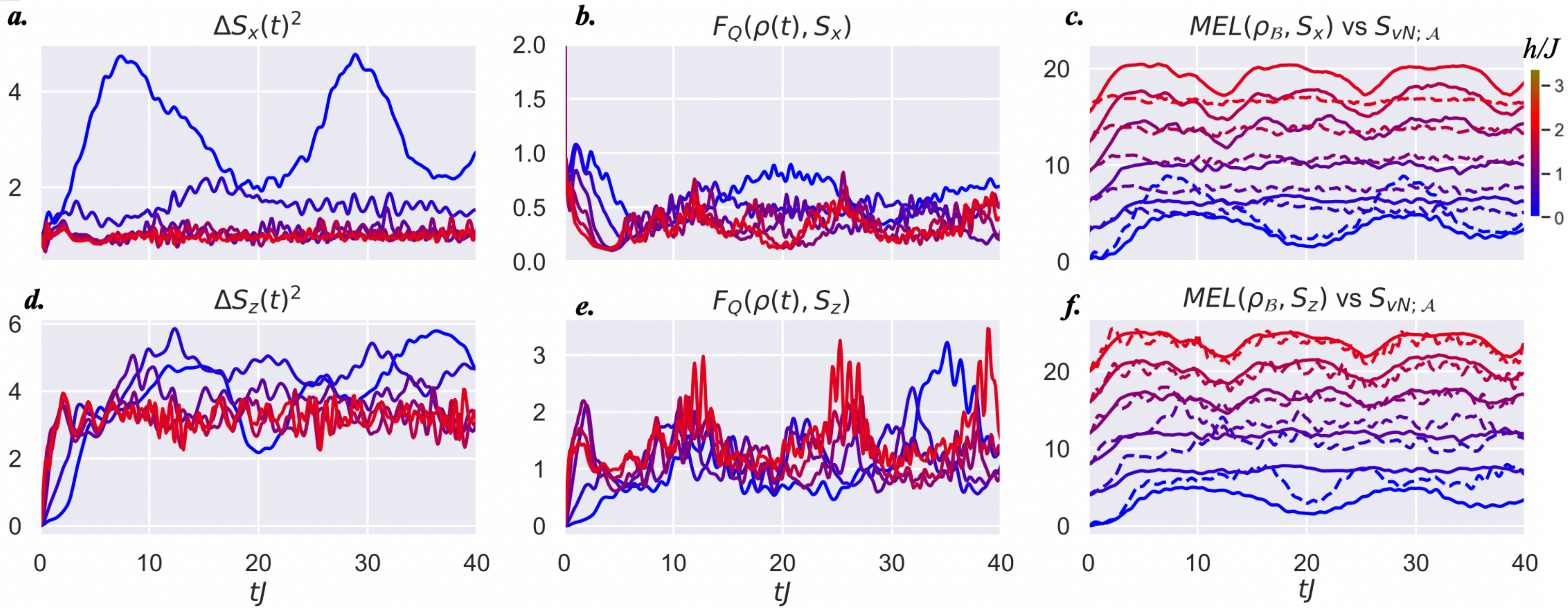}
    \caption{\textbf{MEL dynamics from non-equilibrium quench $(\lambda^2/J\omega_c = 2.0)$}. Real-time evolution of the transverse and longitudinal (top, bottom) spin-fluctuations (a,d), $\mathcal{F}_Q$ (b,e), and multipartite entanglement loss (MEL) (dotted) vs ancilla $S_{vN}$ (solid) (c,f) from polarized initial state. Rapid spin fluctuations are suppressed on the order of ancilla entanglement profile in both $x,z$ spin components. Fisher information is reduced compared to the upper bound set by the fluctuations along the two spin vectors. (c,f) Taking the difference between fluctuations and QFI provides the $MEL$. $MEL(S_x)$ provides a good estimate near $J=0$ (blue) and is most accurate near the minima in $S_{vN;\mathcal{A}}$. About and across the DQPT, $MEL(S_x)$, saturates and does not capture $S_{vN; \mathcal{A}}$. $MEL(S_z)$ always upper bounds $S_{vN; \mathcal{A}}$ for $h/J < 1$ and most accurately captures the entropy at the entropy maxima. $MEL(S_z)$ nearly exactly traces $S_{vN; \mathcal{A}}$ for $h/J > 1$. MEL and $S_{vN}$ curves are offset vertically for clarity. System size $L = 8, d = 40$ and $h$ values $(0, 0.4, 0.8, 1.2, 1.6, 2.0)$.}
    \label{fig:neq_MEL_L}
\end{figure}

\begin{figure}
    \centering
    \includegraphics[width=0.95\textwidth]{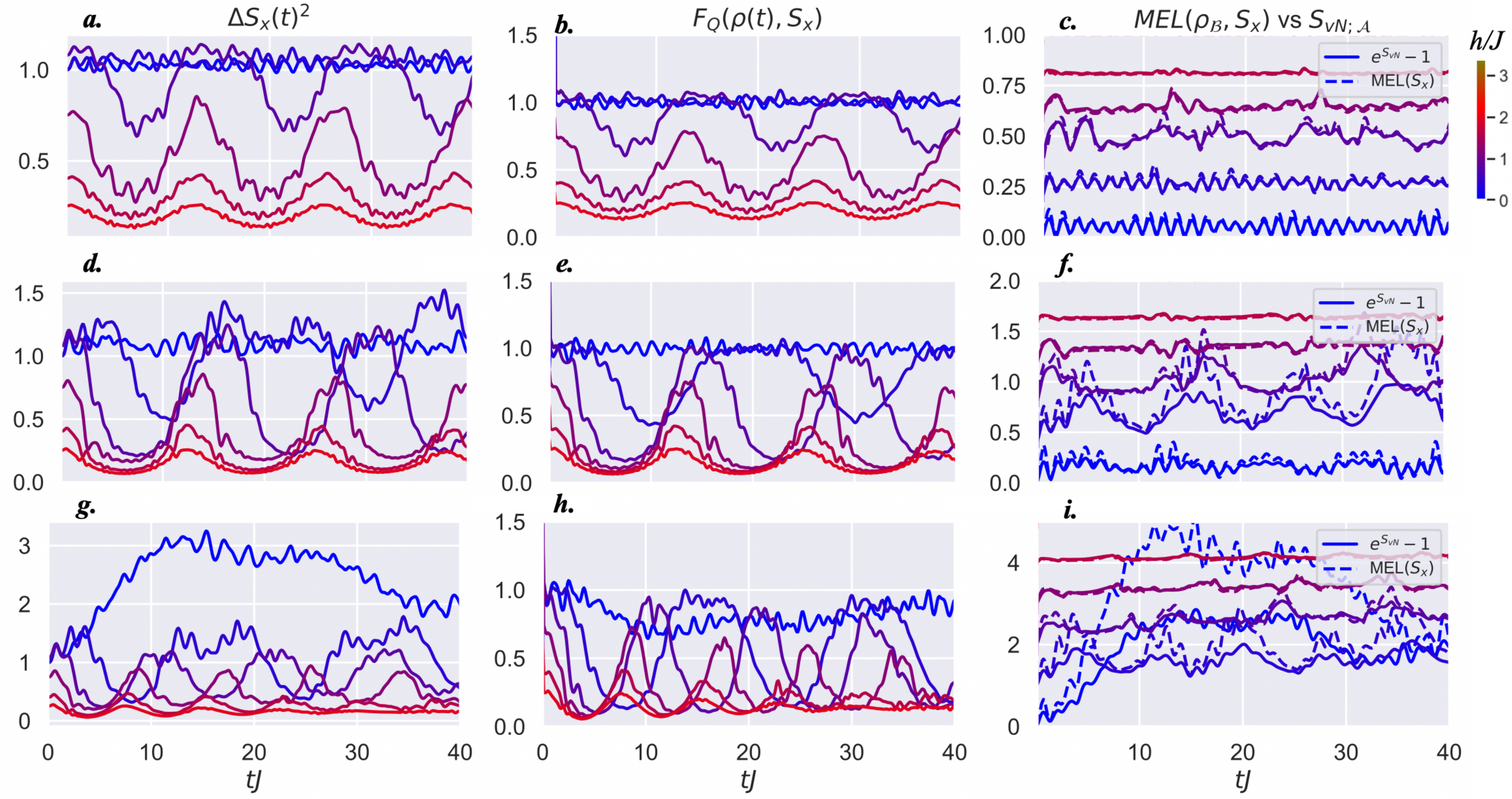}
    \caption{\textbf{MEL dynamics following a quench from the Ising ground state $J=1.0,h/J$}: (a, d, g) transverse spin-fluctuations $\Delta S_x^2$; (b, e, h) Fisher information $\mathcal{F}_Q$; (c, f, i) multipartite entanglement loss (MEL) (dotted) vs ancilla $S_{vN: \mathcal{A}}$ (solid) for $L=8, d=40, \lambda^/\omega_c = 0.18, 0.5, 1.125$ (top, middle, bottom row). Spin fluctuations are suppressed on the order of ancilla entanglement profile, where Fisher information is reduced compared to the upper bound set by the fluctuations. (c,f,i) Taking the difference between fluctuations and information provides MEL an accurate estimate of the ancilla entanglement entropy in all regimes. This breaks down near the critical point and with increasing $\lambda$. MEL and $S_{vN}$ curves are offset vertically for clarity. System size $L=8, d = 40$ and $h$ values $(0, 0.4, 0.8, 1.2, 1.6, 2.0)$.}
    \label{fig:eq_MEL_GS}
\end{figure}

\begin{figure}
    \centering
    \includegraphics[width=0.375\textwidth]{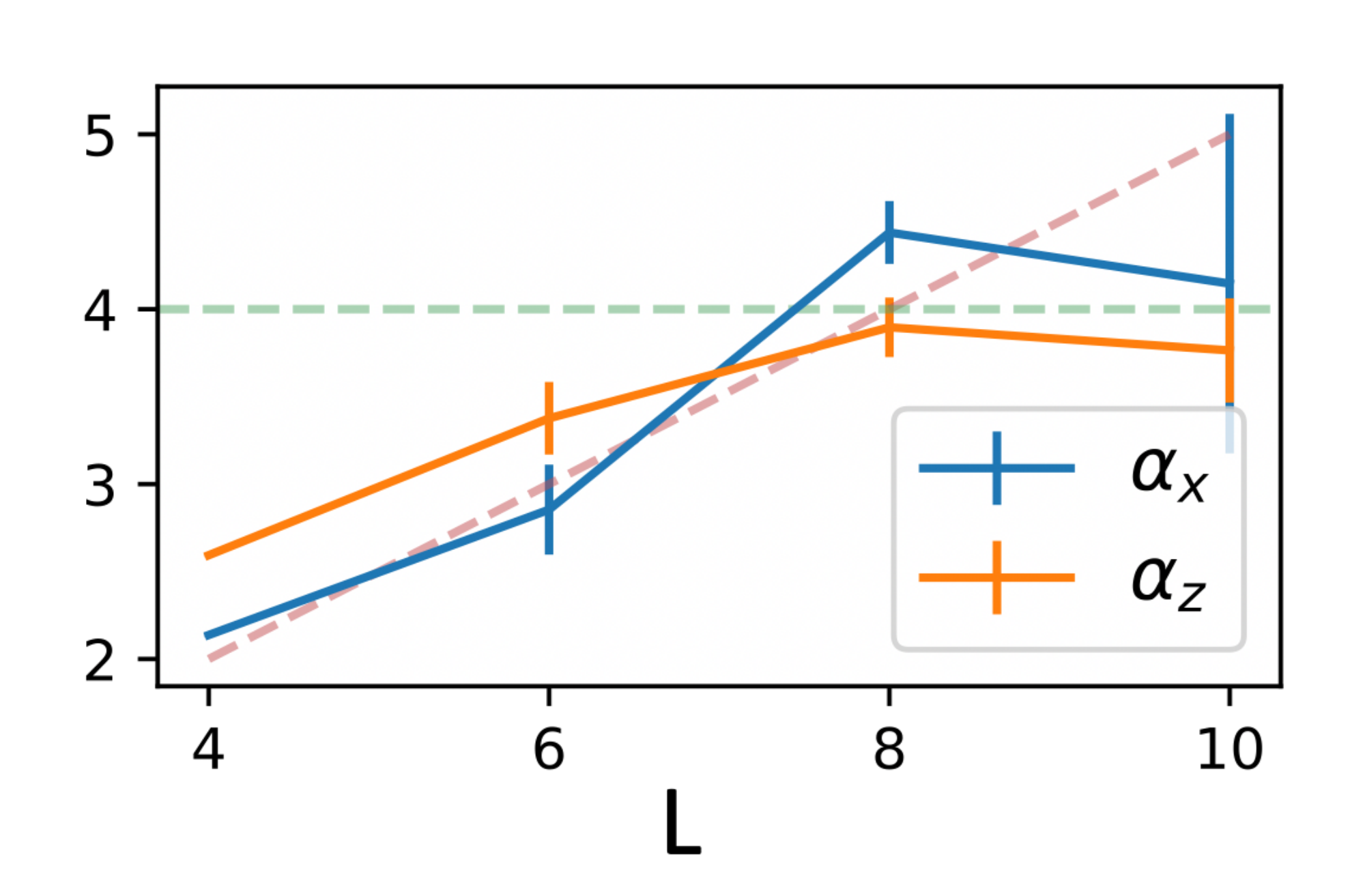}
    \caption{\textbf{Finite size scaling of MEL and entanglement proportionality}: We calculate the long time average of MEL and $S_{vN;\mathcal{A}}$ and determine the constant of proportionality between the two in regimes where they qualitatively agree: MEL$(S_x, \lambda << 1.0)$, MEL$(S_z, h/J \geq 1.0)$. Outside of these regimes  MEL measures along these vectors do not accurately capture the entanglement profile and the general spin-vector of maximal fluctuations is not known. For small systems, the proportionality $\alpha$ grows with system size $\alpha_\mu = \frac{1}{2}L$ until $L = 8$, where the system in the infinite size limit is $4-$partite entangled. $\alpha_x$ and $\alpha_z$ are in strong qualitative agreement and provide nearly identical relations to $S_{vN;\mathcal{A}}$ for all sizes. We attribute the size dependence for small systems on finite size effects that are similarly observed in tandem with the Fisher information density $f_Q$. $f_Q$ grows with system size to the infinite system size limit $f_Q \sim 3$ at roughly $L = 8$ as seen in the main text and large system results provided in ~\cite{Pappalardi_2017}.}
    \label{fig:fs}
\end{figure}

\begin{figure}
    \centering
    \includegraphics[width=0.85\textwidth]{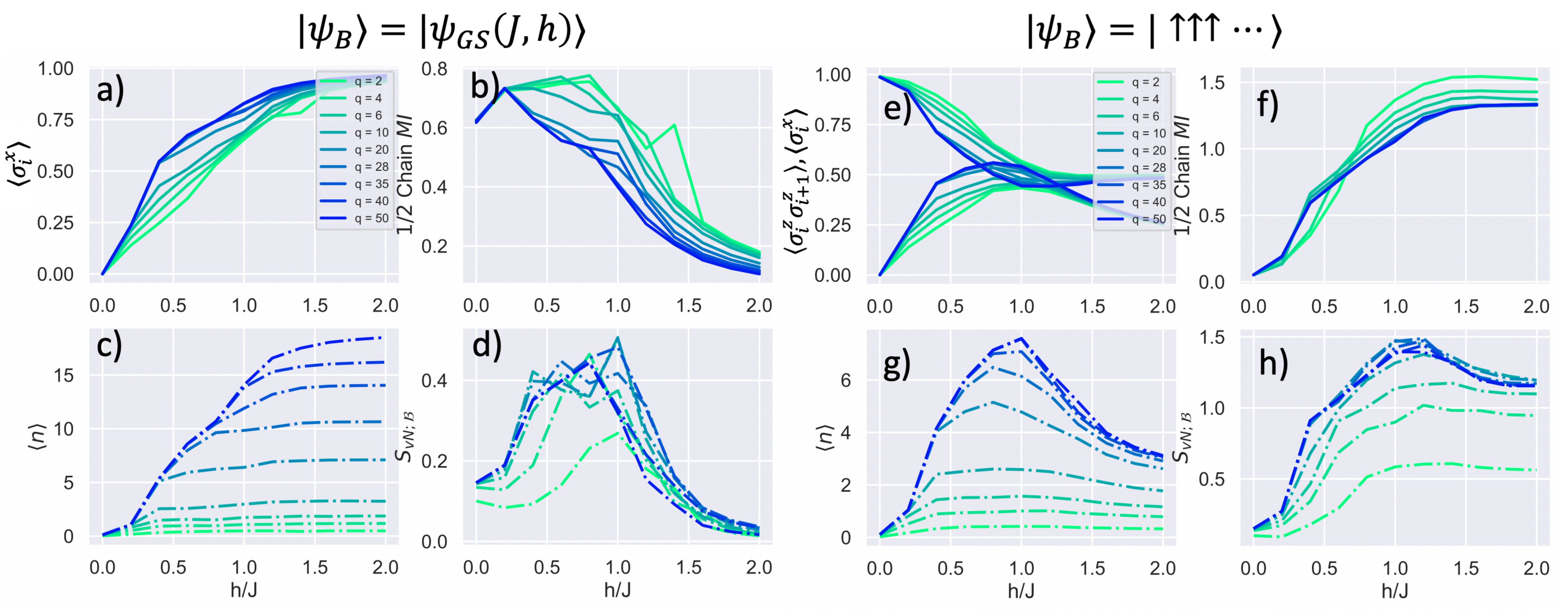}
    \caption{Long time average features following a quench from the Ising ground state (left) and polarized non-equilibrium state (right). The spin-ancilla system is prepared in a product state at $t=0$ and quenched to regions of the Ising phase diagram $J=-1, h/J$ with an additional ancilla-spin coupling $\lambda$. Here $\lambda^2/\omega_c = 0.72$, $L = 12$. (a,e) local transverse magnetization $\langle \sigma^x_i \rangle$ and (e) nearest neighbor correlations $\langle \sigma^z_i \sigma^z_{i+1}\rangle$, (b,f) the $1/2$ chain mutual information, (c,g) the cavity occupation $\langle n \rangle$, and (d,h) the von Neumann entanglement entropy between the qudit and spin-chain. Nearest-neighbor correlations in (e) agree with t-DMRG results presented in ~\cite{DPT_Gorshkov_2019} Results are averaged over the window $tJ = [0,50]$.}
    \label{fig:qVar}
\end{figure}

\bibliography{references}